\documentclass{aa}  

\usepackage{graphicx}
\usepackage{txfonts}
\usepackage{lscape}
\begin{document}

\title{Barium abundance in the central part of the dSph Fornax galaxy.} 

\author { 
S. M. Andrievsky\inst{1,2}\and 
S. A. Korotin\inst{1,4}\and 
V. Hill\inst{3}\and 
A.V. Zhukova\inst{4}
}

\institute {
Department of Astronomy and Astronomical Observatory, 
Odessa National University, and Isaac Newton Institute of Chile Odessa branch,
Shevchenko Park, 65014 Odessa, Ukraine\\
\email{andrievskii@ukr.net}
\and 
GEPI, Observatoire de Paris, PSL, Research University,  CNRS,  Univ Paris 
Diderot, Sorbonne Paris Cit\'e, Place Jules Janssen, 92195 Meudon, France
\and 
Universit\'e C\^ote d'Azur, Observatoire de la C\^ote d'Azur, CNRS, 
Laboratoire Lagrange, Bd de l'Observatoire, CS 34229, 06304, Nice Cedex 4, 
France
\and 
Crimean Astrophysical Observatory, Nauchny, 298409, Crimea
} 

\date{ }
\abstract{
We revisited barium abundance in a sample of giant stars in the dwarf spheroidal 
Fornax galaxy.
Non-local thermodynamic equilibrium (NLTE) synthesis was used to derive the barium abundance 
from two \ion{Ba}{II} lines.
Our new NLTE result for barium shows that in the range of metallicities from 
--1 to --0.5 the abundance of this element is almost the same as in the stars of 
the Milky Way.
We conclude that the evolution of barium abundance in the dSph Fornax galaxy 
resembles that of the Milky Way at the epoch of the corresponding metallicity 
level.}


\maketitle

\section{Introduction.}

In recent years chemical properties of dwarf galaxies from the Local 
Group have attracted the attention of 
Galactic evolution researchers. Two questions are of particular importance.
First,  was our Galaxy formed as a result of the merging process involving smaller 
stellar systems such as the dwarf galaxy and irregular galaxies? Second, is chemical 
evolution in dwarf galaxies similar to or different from the evolution of our Galaxy 
at the similar metallicities? Many papers were published in the literature on 
these topics attempting to answer these two questions. Initially, formation scenarios
considered two possible options. According to the first scenario (\citealt{Egg62}), 
our Galaxy
was formed from a metal-poor, slowly spinning protogalactic cloud of spherical form and, 
after this cloud collapsed, field halo stars and globular clusters were formed.
Another scenario, proposed by \citet{Sea77}, assumed that the Galaxy formed from the "building blocks" of masses of 
about 10$^{8}$ M$_{\odot}$. Such merging fragments then evolved independently 
of each other. The problem of galaxy formation has not been solved
to this point. For example, based on the chemodynamical analysis of the 
Galactic halo, \citet{Car07} proposed that our Galaxy was formed by continuous 
accretion of separate stellar subsystems or protogalactic clumps (i.e., dwarf spheroidal galaxies and 
irregular  galaxies). In contrast, \citet{Gei07} argued that stellar subsystems are rather different from the Galactic halo in their chemical properties and thus the 
accretion scenario is hardly valid.

It is considered to be well established today that dwarf galaxies from the 
Local Group demonstrate a decreased abundance of $\alpha$-elements compared 
to abundances in the Milky Way at the same metallicities. This is explained 
as a result of a different star formation history in dwarf galaxies; i.e.,  there is a lower 
star formation rate in the time-delay model (see \citealt{Mat14}).    

Abundances of $s$-process elements are also important tracers of Galactic 
evolution. The formation of $s$-process elements occurs as a result of free neutron captures 
by seed nuclei in a medium with a low value of neutron flux. In contrast to this
process, the $r$-process produces higher mass nuclei in a medium with a high value of 
neutron flux. The interiors of AGB stars are appropriate places for such processes to operate.  Among the elements that have a large contribution from $s$-process are Sr, Y, Zr, Ba, and La  and these have become the subject of many observational 
and theoretical studies. A high value of barium 
abundance indicates strong  $s$-process production in dwarf galaxies from low-mass stars 
with very long lifetimes. 

The dwarf spheroidal galaxy Fornax is one of the most luminous satellites of the 
Milky Way. Its stellar population was studied spectroscopically by many 
authors. In particular, \citet{Let10} and \citet{Lem14} determined
abundances of $\alpha$- and $s$-process elements in the giants of this stellar 
system. In the first paper, these authors reported about a remarkable overabundance 
of barium in Fornax. In the second paper, the barium abundance results
were revisited, but nevertheless, the new mean [Ba/Fe] value appeared to be rather high as well.

As a rule, the \ion{Ba}{II} lines are used as a barium abundance indicator. Since the rather bright giant stars in external galaxies are 
more readily available for high resolution spectroscopy, one should expect 
that the \ion{Ba}{II} lines seen in their spectra are strong. The abundance analysis of these 
strong lines require some additional attention and special methods. The abundances of $s$-process elements in the 
stars of satellite dwarf galaxies are of  great importance both for 
understanding their evolution and the evolution of our Galaxy. Therefore, we decided to 
reconsider barium abundance in the dwarf Sph Fornax galaxy without resorting to the LTE 
approximation.
\section{Stellar sample and method}

Our program star sample was previously investigated by 
\citet{Let10}. The spectra were secured with ESO VLT facilities with 
resolving power of about 20000 in the two ranges 5340--5620 \AA~and 
6120--6701 \AA. Eighty-one RGB stars were selected in the central part of the Fornax dSph 
galaxy. In contrast to the above-mentioned work, where authors derived 
abundances of 12 chemical elements, we only concentrated on one 
element, i.e., barium, because Letarte et al. reported a puzzling
strong overabundance of this element in the Fornax giants. 

We used stellar parameters and iron abundances determined by \citet{Let10}. 
The program stars and their iron abundances are listed in Table \ref{tabab} 
together with our NLTE abundance of barium.
                             
The barium abundance for each star was determined without the LTE approximation. 
For our NLTE calculations we used MULTI code (Carlsson 1986) later modified 
by Korotin et al. (1999). Our barium atomic model contains 31 levels of \ion{Ba}{I}, 
101 levels of \ion{Ba}{II} with quantum number $n$ < 50, and  the ground level of \ion{Ba}{III} 
ion. We computed in detail 91 bound-bound transitions between the first 28 levels of \ion{Ba}{II} ($n$ < 12 and $l$ < 5). 
Information about the adopted oscillator strengths,
photoionization cross-sections, collisional rates, and broadening parameters can be found 
in (\citealt{And09}). To derive the barium abundance, we fitted the NLTE synthetic profiles with the observed profiles of two \ion{Ba}{II} lines, i.e., 6141 and 6496 \AA.

Within the considered range of atmosphere parameters, barium lines do not 
show large deviations from LTE. At the same time NLTE corrections demonstrate 
complex behavior that depends on atmosphere parameters and barium abundance itself 
\citep{Kor15}. For the 6496 \AA~ barium line, the hyperfine structure ({\it hfs}) was taken 
into account, albeit its influence on the line profile is negligible. The distance between 
{\it hfs} components is less than 0.012 \AA.

Because of the modest resolving power of our spectra (R=20000) the barium 
line 6496~\AA~is partially blended in its blue wing with neutral iron line 
6496.4~\AA. The line at 6141~\AA\  is also blended with the line of the neutral iron 
at 6141.7~\AA, which  means that it is not possible to measure the equivalent widths of these lines accurately. Therefore we investigated their profiles 
using the synthetic spectrum technique. 

\begin{figure}
\resizebox{\hsize}{!}                  
{\includegraphics {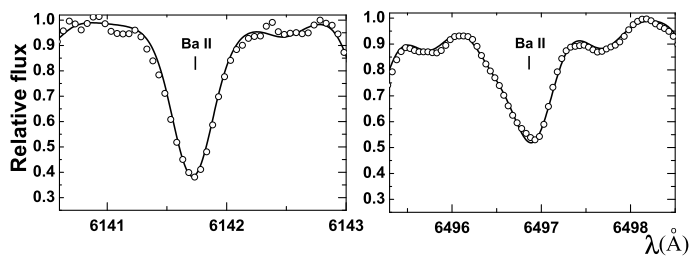}}
\caption[]{Synthetic and observed spectra of two barium lines for the star BL210. The 
best barium abundance is (Ba/H)=1.67, here (Ba/H)=$\log\epsilon(\rm Ba)$+12.00). 
Circles indicate the observed spectrum and the continuous line indicates the synthetic spectrum.}
\label {baprof}
\end{figure}

As  in our paper series on NLTE analysis, we used a 
combination of the NLTE calculation of the $b$-factors (the ratio between NLTE 
and LTE atomic level populations) and the LTE spectrum synthesis (SYNTHV code; 
\citealt{Tsy96}). With the latter code we generated synthetic spectrum in 
the vicinity of each barium line taking into account all the lines in the 
considered wavelength region listed in the Vienna Atomic Line Database (VALD;  
\citealt{Piset95}, \citealt{Kup99})
\footnote{http://vald.astro.univie.ac.at/~vald3/php/vald.php}. 
While those lines were treated in LTE, for the profile synthesis 
of the barium lines we used previously calculated $b$-factors. We included these factors in 
SYNTHV to calculate the NLTE barium line source function, and using this 
procedure we were able to take into account the lines of other species that may 
blend the barium lines of interest. An example of the NLTE synthetic profiles 
fitted to observed profiles is shown in Fig. \ref{baprof}.  

Table \ref{hfs} contains information about barium line parameters 
used in our calculations.

\begin {table}
\caption {Barium line parameters.}
\label {hfs}
\begin{tabular}{l c c  c }
\hline
Ion &$ \lambda$ (\AA) & {\it hfs} & log gf \\
  &           &$\Delta\lambda$, m\AA&           \\
\hline
\hline
\ion{Ba}{II} & 6141.71&   -&  --0.08 \\
\hline
\ion{Ba}{II} & 6496.90&   0&  --0.46 \\
\ion{Ba}{II} &        &    --4&  --1.32 \\
\ion{Ba}{II} &        &       8&  --1.55 \\ 
\hline
\hline  
\end {tabular}  
\end {table}

\section{Results and discussion}

It is generally accepted that barium nuclei are produced in the $r$-process at 
early stages of Galactic evolution and later in the $s$-process in lower mass AGB stars along with even heavier $s$-process elements 
\citep[e.g.,][]{Bus99}. Among other factors, the efficiency of barium nuclei production also depends on the metallicity of AGB stars. Metal-poor AGB 
stars produce a larger amount of second $s$-process peak nuclei. An
increased ratio [Ba/Fe] in an external galaxy can either be a sign of the peculiar
efficiency of barium production by AGB stars of that stellar system or  the
existence of an additional source of heavy $s$-process nuclei. 

\citet{Let10} performed the LTE determination of the barium 
abundance in Fornax galaxy. For the program stars
in the 
metallicity range from approximately --1 to --0.5, these authors obtained a remarkably high 
barium abundance when compared with, for example, the Milky Way data, i.e., [Ba/Fe] = 
+0.62 $\pm 0.29$  (see their Fig. 14). Later, \citet{Lem14} added 
47 Fornax red giant branch stars to their
analysis. New data on barium abundance from 
\citet{Lem14} gave the mean [Ba/Fe] value +0.37 $\pm 0.17$. In their Fig. 15 these authors 
also plotted reconsidered barium abundance data from \citet{Let10} and, after 
reconsidering, gave almost the same mean barium abundance value. Our mean barium 
abundance in this metallicity domain is +0.13 $\pm 0.17$, which is much lower than 
reported before by the above-mentioned authors. 

\begin{figure}
\resizebox{\hsize}{!}
{\includegraphics {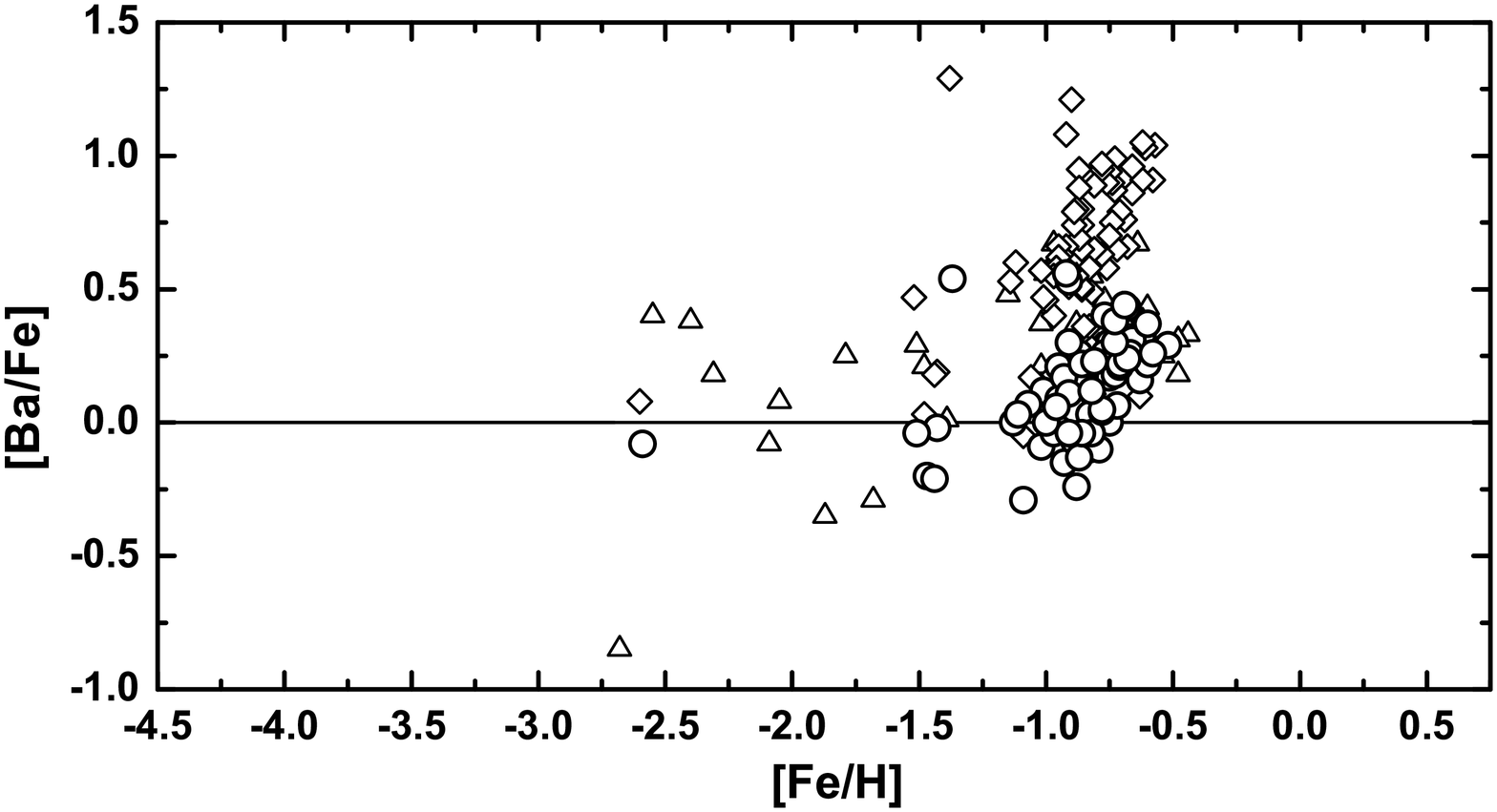}}
\resizebox{\hsize}{!}
{\includegraphics {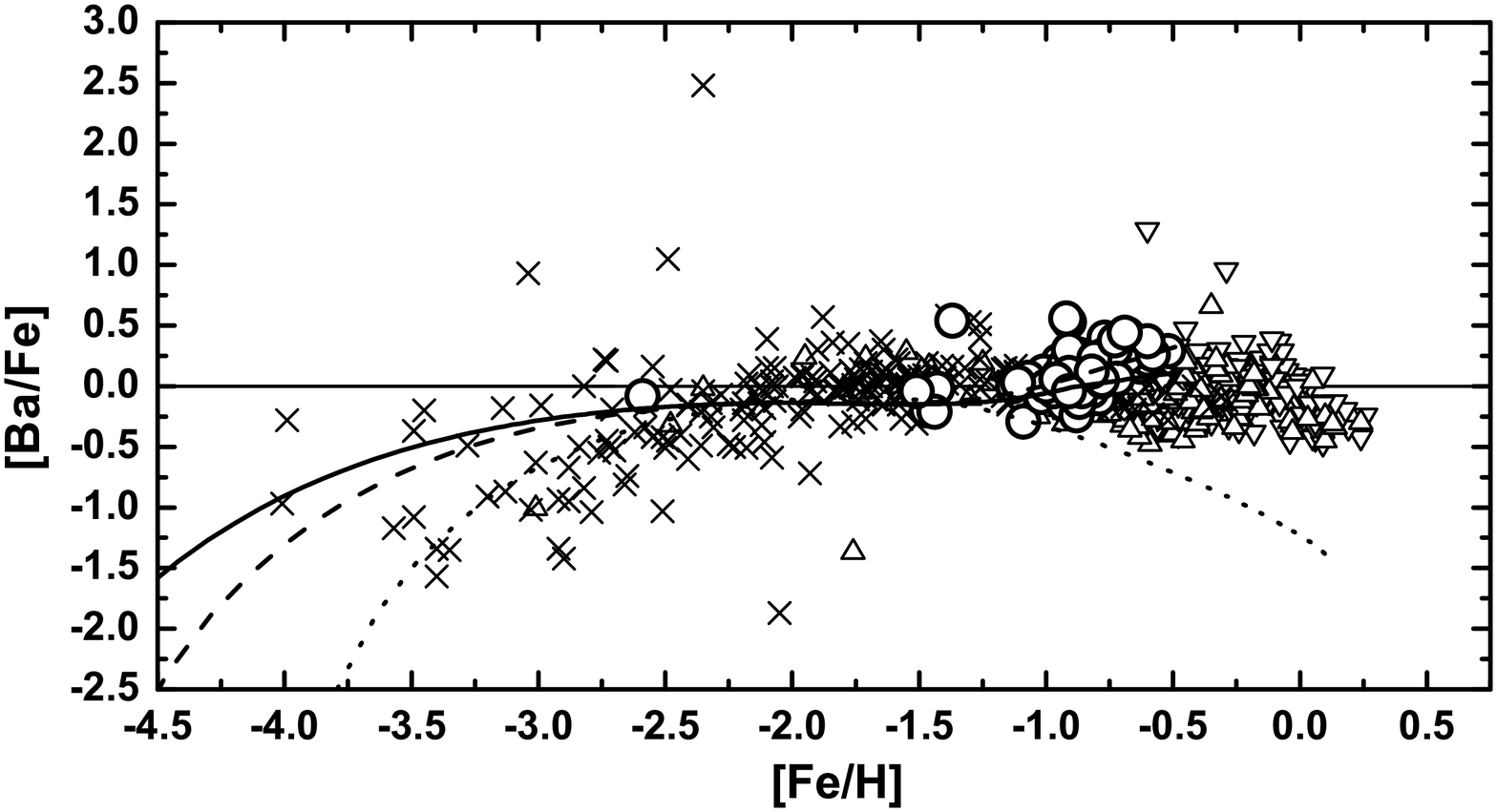}}
\caption[]{[Ba/Fe] vs. [Fe/H].   $Upper~panel$ -- Our barium abundance data are shown ($open~circles$)
together with barium abundance data derived by \citet{Let10} ($open~diamonds$) and \citet{Lem14}
($open~triangles$). $Lower~panel$ -- Milky Way stars are shown. \citet{Ven04} compilation: $triangles$ indicate 
thin and thick disk stars, $crosses$ indicate halo stars, and open circles indicate our program stars from 
the Fornax galaxy.  
The \citet{Bis17} model prediction for the Milky Way: the $thick~line$ indicates a thin disk, the $dashed~line$ indicates a thick disk,  and the $dotted~line$ indicates a halo.}
\label {bagal}
\end{figure}

In Fig. \ref{bagal} we present our individual barium abundances in 
Fornax stars graphically together with \citet{Let10}, \citet{Lem14}, and the Milky Way data, indicating 
the positions of the stars from thin and thick disks and also the Galactic halo. Such a comparison can be affected by intrinsic offsets since the 
Milky Way barium abundance data were derived in LTE, while we show the NLTE results 
for our program stars. Nevertheless, the NLTE corrections for the barium abundance at 
this metallicity, temperature, and gravity regime are very small \citep[see][]{Kor15}, 
and thus, one can note that [Ba/Fe] ratios in our Fornax program stars are close to 
typical values for the Milky Way stars in the above-mentioned metallicity region.

In Fig. \ref{bagal} we also plotted the theoretical model prediction for 
the Milky Way provided by \citet{Bis17}. Theoretical curves for the thin and thick disk
stars in the Galaxy describe positions for our Fornax stars very well.

\section{Conclusions}

We derived barium abundance for a sample of giant stars from the central part of the 
dSp Fornax galaxy. The barium abundance in 81 program stars was determined using the
sophisticated NLTE approximation. Previous determinations (\citealt{Let10} and \citealt{Lem14}) have
led to the conclusion that barium abundances are enhanced in these stars. According to 
\citet{Redlam17},  the enhanced barium abundance often reported for solar twins, stellar 
associations, and  open cluster stars at young ages is unrelated to stellar nucleosynthesis 
peculiarities, but on the contrary it results from an overestimation of barium abundance obtained 
through standard methods of LTE abundance analysis. Our NLTE spectroscopic analysis 
reveals  that our program stars from the central part of the dSph Fornax galaxy do not display 
significantly increased barium abundances compared to the Milky Way stars. This may indicate 
that no special mechanism is required to explain the high LTE barium abundance in the Fornax galaxy 
found in previous studies. In this sense, the evolution of barium abundance in the dSph Fornax 
galaxy resembles its evolution in the Milky Way at the epoch of the corresponding metallicity level.

\begin{acknowledgements}
We are thankful to Dr. Sara Bisterzo for her kind help with theoretical modeling data.
SMA is thankful to the Universite C\^ote d'Azur, Observatoire de la C\^ote d'Azur, CNRS, 
Laboratoire Lagrange administration for their hospitality and financial 
support during his visit
in 2016, and also for the partial financial support from the SCOPES grant No. IZ73Z0-152485, 
which is also acknowledged by SAK. We also thankful to the referee for many
valuable comments that improved our paper.
\end{acknowledgements}

\begin{appendix}
\section{Program stars, their metallicity, and abundances of Ba.}

\begin{table*}
\caption{Program stars, their metallicity, and abundances of Ba; (Ba/H)=$\log\epsilon(\rm Ba)$+12.00; [Ba/H] = (Ba/H)$_{\rm star}$ -- (Ba/H)$_{\rm Sun}$.}
\label{tabab}
\begin{center}  
\begin{tabular}{lccc|lccc}
\hline                                                     
\hline                                                     
  Star    &  [Fe/H]& (Ba/H) &[Ba/Fe] &  Star    &  [Fe/H]& (Ba/H) &[Ba/Fe]\\
\hline                                                     
  BL038   & --0.88 &   1.32  &  0.03   & BL185   & --0.71  & 1.67  &  0.21 \\
  BL045   & --1.09 &   0.79  &--0.29   & BL190   & --0.79  & 1.28  &--0.10 \\
  BL052   & --1.02 &   1.06  &--0.09   & BL195   & --0.97  & 1.16  &--0.04 \\
  BL065   & --1.43 &   0.72  &--0.02   & BL196   & --1.07  & 1.17  &  0.07 \\
  BL076   & --0.85 &   1.43  &  0.11   & BL197   & --0.89  & 1.33  &  0.05 \\
  BL077   & --0.79 &   1.43  &  0.05   & BL203   & --0.83  & 1.37  &  0.03 \\
  BL079   & --0.52 &   1.94  &  0.29   & BL204   & --1.00  & 1.17  &  0.00 \\
  BL081   & --0.62 &   1.88  &  0.33   & BL205   & --0.69  & 1.78  &  0.30 \\
  BL084   & --0.85 &   1.44  &  0.12   & BL208   & --0.66  & 1.82  &  0.31 \\
  BL085   & --2.59 &  --0.50 &--0.08   & BL210   & --0.76  & 1.67  &  0.26 \\
  BL091   & --0.96 &   1.25  &  0.04   & BL211   & --0.67  & 1.76  &  0.26 \\
  BL092   & --0.95 &   1.25  &  0.03   & BL213   & --0.93  & 1.41  &  0.17 \\
  BL096   & --0.75 &   1.42  &  0.00   & BL216   & --0.77  & 1.66  &  0.26 \\
  BL097   & --0.92 &   1.41  &  0.16   & BL218   & --0.60  & 1.79  &  0.22 \\
  BL100   & --0.93 &   1.09  &--0.15   & BL221   & --0.86  & 1.53  &  0.22 \\
  BL104   & --0.96 &   1.29  &  0.08   & BL227   & --0.91  & 1.56  &  0.30 \\
  BL113   & --0.75 &   1.59  &  0.17   & BL228   & --0.88  & 1.05  &--0.24 \\
  BL115   & --1.47 &   0.50  &--0.20   & BL229   & --0.71  & 1.68  &  0.22 \\
  BL123   & --0.97 &   1.20  &  0.00   & BL233   & --0.68  & 1.73  &  0.24 \\
  BL125   & --0.73 &   1.62  &  0.18   & BL239   & --0.91  & 1.37  &  0.11 \\
  BL132   & --0.89 &   1.21  &--0.07   & BL242   & --1.11  & 1.09  &  0.03 \\
  BL135   & --0.95 &   1.43  &  0.21   & BL247   & --0.82  & 1.31  &--0.04 \\
  BL138   & --1.01 &   1.28  &  0.12   & BL250   & --0.68  & 1.92  &  0.43 \\
  BL140   & --0.87 &   1.44  &  0.14   & BL253   & --0.73  & 1.74  &  0.30 \\
  BL141   & --0.82 &   1.31  &--0.04   & BL257   & --0.58  & 1.85  &  0.26 \\
  BL146   & --0.92 &   1.35  &  0.10   & BL258   & --0.60  & 1.94  &  0.37 \\
  BL147   & --1.37 &   1.34  &  0.54   & BL260   & --0.87  & 1.17  &--0.13 \\
  BL148   & --0.63 &   1.70  &  0.16   & BL261   & --0.86  & 1.27  &--0.04 \\
  BL149   & --0.91 &   1.34  &  0.08   & BL262   & --0.78  & 1.43  &  0.04 \\
  BL150   & --0.83 &   1.24  &--0.10   & BL266   & --1.44  & 0.52  &--0.21 \\
  BL151   & --0.86 &   1.41  &  0.10   & BL267   & --0.72  & 1.51  &  0.06 \\
  BL155   & --0.75 &   1.72  &  0.30   & BL269   & --0.81  & 1.59  &  0.23 \\
  BL156   & --1.13 &   1.04  &  0.00   & BL278   & --0.73  & 1.82  &  0.38 \\
  BL158   & --0.87 &   1.56  &  0.26   & BL279   & --1.51  & 0.62  &--0.04 \\
  BL160   & --0.95 &   1.31  &  0.09   & BL295   & --0.69  & 1.92  &  0.44 \\
  BL163   & --0.77 &   1.80  &  0.40   & BL300   & --0.92  & 1.81  &  0.56 \\
  BL166   & --0.89 &   1.44  &  0.16   & BL304   & --0.96  & 1.27  &  0.06 \\
  BL168   & --0.88 &   1.44  &  0.15   & BL311   & --0.78  & 1.44  &  0.05 \\
  BL171   & --0.90 &   1.33  &  0.06   & BL315   & --0.82  & 1.47  &  0.12 \\
  BL173   & --0.85 &   1.33  &  0.01   & BL323   & --0.91  & 1.22  &--0.04 \\
  BL180   & --0.91 &   1.79  &  0.53   &         &         &       &       \\
\hline                                                     
\end{tabular}                                              
\end{center}    
\end{table*}

\end{appendix}

\end{document}